\documentclass[a4paper,11pt]{article}
\usepackage{aaskaiid}
\usepackage{orcidlink}
\usepackage[version=4]{mhchem}
\usepackage{xspace}
\newcommand{\arcsec}{\ensuremath{^{\prime\prime}}\xspace}

\newcommand{\AAstar}{AA$^{*}$\xspace}

\title{SKAO--ALMA Synergies in Star Formation Science}
\ShortTitle{SKAO--ALMA Synergies}

\author[1]{Jan Forbrich\orcidlink{0000-0001-8694-4966}}
\ShortName{J. Forbrich et al.} 
\author[2]{Gary A. Fuller\orcidlink{0000-0001-8509-1818}}
\author[3]{Mark A. Thompson\orcidlink{0000-0001-5392-909X}}
\author[4]{Eleonora Bianchi\orcidlink{0000-0001-9249-7082}}
\author[5]{Jaime E. Pineda\orcidlink{0000-0002-3972-1978}}
\author[6,4]{Marta De Simone\orcidlink{0000-0001-5659-0140}}
\author[7]{Tomoya Hirota\orcidlink{0000-0003-1659-095X}}
\author[8]{Katharine Johnston}
\author[9]{Steven N. Longmore}
\author[4]{Giovanni Sabatini\orcidlink{0000-0002-6428-9806}}
\affiliation[1]{University of Hertfordshire, Centre for Astrophysics Research, College Lane, Hatfield AL10 9AB, UK}
\emailAdd{j.forbrich@herts.ac.uk}
\affiliation[2]{Jodrell Bank Centre for Astrophysics, School of Physics and Astronomy, University of Manchester, Oxford Road, Manchester, M13 9PL UK}
\affiliation[3]{School of Physics and Astronomy, University of Leeds, Leeds LS2 9JT, UK}
\affiliation[4]{INAF, Osservatorio Astrofisico di Arcetri, Largo E. Fermi 5, I-50125, Firenze, Italy}
\affiliation[5]{Max-Planck-Institut f{\"u}r Extraterrestrische Physik, Giessenbachstr. 1, D-85748 Garching bei M{\"u}nchen}
\affiliation[6]{European Southern Observatory, Karl-Schwarzschild-Strasse 2, 85748, Garching bei Munchen, Germany}
\affiliation[7]{Mizusawa VLBI Observatory, National Astronomical Observatory of Japan, 2-12 Hoshigaoka-cho, Mizusawa, Oshu-shi, Iwate 023-0861, Japan; SOKENDAI (The Graduate University for Advanced Studies), 2-21-1 Osawa, Mitaka-shi, Tokyo, 181-8588, Japan}
\affiliation[8]{School of Engineering and Physical Sciences, Isaac Newton Building, University of Lincoln, Brayford Pool, Lincoln LN6 7TS, UK}
\affiliation[9]{Astrophysics Research Institute, Liverpool John Moores University, 146 Brownlow Hill, Liverpool L3 5RF, UK}

\abstract{We highlight the potential for synergies between the SKA telescopes and ALMA, which will begin to be significant already with SKA Early Science. Broadly within star formation science, we focus on 1) (simultaneous) time domain and variability studies (e.g., of (proto-)stars), which is systematically opening a new observing window that will enable us to better constrain the physics of extreme stellar flares and their interplay with mass accretion in young stars, 2) possibilities for complementary coverage of different transitions of molecules as tracers of the ISM (e.g., \ce{H2CO} to trace density and temperature), and 3) radio recombination lines to trace ionised gas at different densities. Furthermore, we address synergies in terms of the spatial resolution of both observatories and new capabilities not only on the SKAO side but extending to the ALMA2030 wideband sensitivity upgrade.}

\begin{document}
\maketitle

\section{Context: resolved molecular clouds and young stellar objects}

The advent of the Square Kilometre Array Observatory (SKAO) coincides with the routine availability of the Atacama Large Millimeter Array (ALMA), both located in the Southern hemisphere. Additionally, the advent of the full design baseline of the SKA telescopes in the 2030s coincides with that of the improved capabilities of the ALMA2030 wideband sensitivity upgrade. As a result, unprecedented synergies from the operation of two powerful radio interferometers in the Southern hemisphere will arise, covering a wide range of radio frequencies. 

This chapter will discuss the impact of these synergies on star formation science, in both continuum and spectral line observations. We address observations of both the initial conditions of star formation in molecular clouds and also observations of Young Stellar Objects across the full mass range.

\section{The basics: relevant SKAO and ALMA capabilities}

Both ALMA and SKAO are located in the Southern hemisphere, with ALMA at a latitude of $-23^\circ$, compared with SKA-Low at almost $-27^\circ$ and SKA-Mid at almost $-31^\circ$. 
This chapter thus naturally focuses on observations of the Southern sky, with detailed considerations of simultaneous source coverage discussed in the subsection on time domain science.

For the SKAO, we consider the full design baseline (AA4) and early science capabilities (\AAstar). The AA4 design baseline consists of 512 stations for SKA-Low and 197 dishes for SKA-Mid. In contrast, the \AAstar setup for early science consists of 307 SKA-Low stations and 144 SKA-Mid dishes. 
SKA-Low will cover frequencies 50--350~MHz, and SKA-Mid will cover frequencies 350~MHz--15.4~GHz, with four initial frequency bands. For SKA-Low, the typical continuum synthesised beam size will be about 5\arcsec  (both \AAstar and AA4), and in the case of SKA-Mid, the continuum synthesised beam size will initially (\AAstar) range from about one arcsecond (Band~1, 0.8~GHz) to about 0.1\arcsec at 12~GHz (Band~5b). With the advent of AA4, this will improve to 0.3\arcsec and 0.02\arcsec, respectively.

For ALMA, we consider the current observational status, including the newly available Band~1 (35--50~GHz) at the low-frequency end of ALMA operations and up to ALMA Band 10 at 950~GHz. Furthermore, given the SKAO timeline, we include considerations of the ALMA2030 Wideband Sensitivity Upgrade (WSU), which will quadruple the instantaneous bandwidth from the current 8~GHz per polarization. Together with other improvements, this will lead to better sensitivity, primarily in continuum but also in spectral line observations (in addition to the improved instantaneous spectral coverage). Overall, across its ten observing bands and its ten different array configurations, the main ALMA array (12-m dishes) achieves synthesised beam sizes of between 0.1\arcsec and 8\arcsec in Band~1 (40~GHz) and between 5~mas and 0.4\arcsec in Band~10 (870~GHz).

Other than obvious complementarity in terms of spectral coverage of the radio and sub-millimetre bands at broadly comparable spatial sensitivity, there are also crucial differences between the two observatories -- most notably in terms of efficient survey capabilities and commensal operations, both of which are primary design considerations for the SKA telescopes.

\section{Continuum science}

\begin{figure}[ht]
    \centering
	\includegraphics[width=0.50\columnwidth]{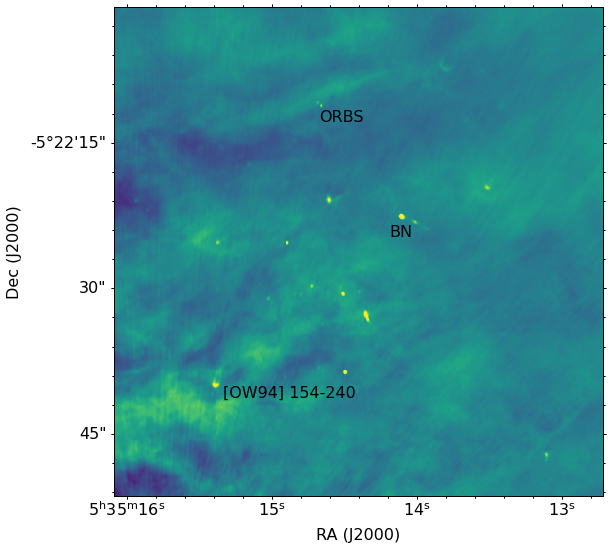}
    \raisebox{15pt}{\includegraphics[width=0.415\columnwidth]{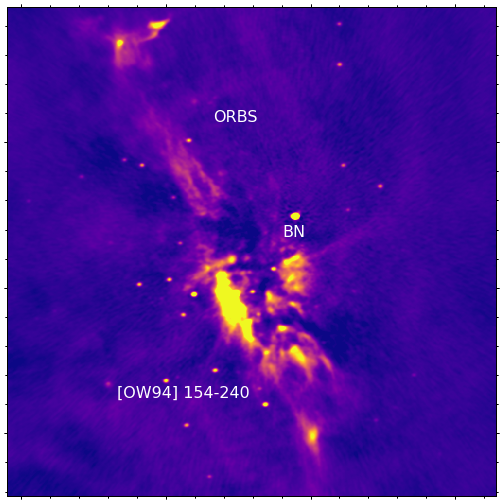}}
    \caption{The BN/KL region in Orion as an example of two continuum views {\it in identical fields}: {\it left:} a deep VLA 10~GHz view \citep{for16}, and {\it right:} an ALMA 100~GHz view \citep{var23}. 
    The point source population is clearly complementary, with only partial overlap. Source BN is marked for orientation, as well as deeply embedded young stellar object ORBS (where an extreme cm-radio flare was observed by \citealp{for08} and a mm-radio flare by \citealp{var23}; see Figure~\ref{fig:ori2}) and a famous proplyd, 154-240.}
    \label{fig:ori}
\end{figure}

While both the frequency range covered by ALMA and that covered by the SKA telescopes each cover a wealth of science cases by themselves, we here focus on the synergies where both can be used in conjunction. The primary science case here is the improved ability to disentangle multiple emission processes by their spectral indices, both by a vastly enlarged spectral baseline and also by covering multiple subsets of the overall spectral range. An additional capability lies in the combination of polarimetric observations across the radio range jointly covered by the SKA telescopes and ALMA, but this is covered elsewhere.

There are at least four mechanisms that produce continuum emission across the radio-sub-mm frequency range. These are the interaction of energetic ionised particles (usually electrons) with a magnetic field, the interaction of ionised particles in an ionised gas giving rise to free-free or Bremsstrahlung radiation, and the thermal emission from dust (e.g., \citealp{dul85,gue02} for an overview).  In addition, there can be emission from spinning dust \citep[e.g.,][]{Draine1998-Spinning_Dust,Draine2003-Dust_Review,planck15XXV}. 
Generally speaking, in a Young Stellar Object (YSO), there will be non-thermal emission from coronal-type activity very close to the central object, with thermal emission from ionised material, for example, at the base of outflows slightly further out. Finally, several emission processes, including dust emission, are operating in protoplanetary disks (see \citealp{fme99} for an early review).
The shape of the spectrum resulting from the first of these depends on the energy of the particles giving rise to gyrosynchrotron or synchrotron radiation with increasing electron energy. Observations over broad spectral ranges can constrain the contributions of these physical environments to the spectral energy distributions of sources \citep[e.g.,][]{dul85}.
This is of particular interest in the case of variable sources, for example stellar activity. For the first time, the availability of sensitive observations across the radio range will enable a step toward what is long possible in solar physics: spectral index time series of different electron populations in stellar flares, where the Sun provides evidence for the existence of possibly separate electron populations at the highest energies traced in the (sub)-mm wavelength regime \citep{kau04,kau09}. On larger scales, for entire star-forming regions or even molecular clouds, the interest is less in variability and more in disentangling, for example, the free-free contamination in the millimetre-range that is otherwise probing thermal dust emission.

Figure~\ref{fig:ori} shows an example of how different YSOs in the same region (Orion BN/KL) appear at frequencies of 10~GHz and 100~GHz, with different physical processes playing a role. There is clearly an overlap between the point source populations, but there are also significant differences. Some sources are only visible in the time domain. The highlighted source Orion Radio Burst Source ORBS showed a major radio flare at cm wavelengths as reported by \citet{for08}. While it is not detected in the integrated 100~GHz image shown (with a cumulative on-source time of 9.3 hours), time-sliced imaging revealed that the source shows a significant mm-wavelength flare and is only detectable during the flare, for about 10 minutes \citep{var23}. 

For individual YSOs, centimetric radio emission first of all provides an extinction-free means to detect YSOs. Already in early evolutionary stages, ionised emission from jets can be detected, combined with nonthermal gyrosynchrotron emission from coronal activity, preferentially in later stages (e.g., \citealp{fme99}). While coronae and jets can be disentangled with high-resolution observations for the most nearby sources, the more common challenge is to separate emission components in unresolved observations. In this regard, higher continuum sensitivity will lead to better detections of polarization, and an increased frequency baseline will improve spectral index studies, including when combining SKA telescopes and ALMA.

The combination of the SKA telescopes and ALMA will enable detailed studies of high-energy processes from time domain studies of flares, where emission mechanisms can be disentangled from detailed lightcurves at different radio frequencies. Such studies require simultaneous observations and have presently largely been limited to the joint use of the VLA and ALMA, which is far more restricted in the accessible Declination range. We discuss such observations in Section~\ref{sec_sim}.

The fact that centimetric YSO radio emission is hardly affected by extinction leads to an additional application: phase-referenced observations offer precision astrometry for objects that may be invisible at optical wavelengths  either because of foreground extinction or intrinsic faintness and hence undetectable by {\it Gaia}. Beyond the scientific interest in the physical origin of the emission, this means that it is possible to obtain astrometric time series to measure (absolute) proper motions and to conduct reflex motion to determine orbital motions \citep{Maureira2020-IRAS16293_Orbits,dzi21,2022ApJ...930...91D} and searches for unseen companions.

On the scale of clumps to clouds and for individual high-mass star formation regions in our Galaxy and nearby galaxies, the joint availability of SKA telescopes and ALMA primarily entails the ability to disentangle centimetric free-free bremsstrahlung emission originating in H\,{\sc ii} regions from thermal emission of cold dust in the surrounding molecular material (clouds). This will be possible in the entire Southern sky at unprecedented sensitivity.

\section{Spectral line science} \label{sec:SpecLine}

The synergy between the SKA telescopes and ALMA is manifold in the area of spectral line science. In the following, we separately discuss molecular clouds and individual YSOs.

\subsection{Molecular clouds and interstellar medium}

The most obvious connection lies in the combination of CO as a tracer of molecular clouds with the capability of tracing the emergence of molecular hydrogen by also observing the atomic hydrogen (H\,{\sc i}) in the surroundings of molecular clouds. 
ALMA can additionally characterise the emergence of molecular gas by adding the two atomic carbon fine-structure transitions in the submillimetre wavelength range.

The combination of \ce{CO} and H\,{\sc i} can be used to study the atomic-to-molecular transition in molecular clouds, the H\,{\sc i}-to-\ce{H2} transition. 
The H\,{\sc i}-to-\ce{H2} transition marks where diffuse atomic hydrogen becomes molecular, enabling efficient cooling and the onset of star formation. 
Understanding the environmental factors that control this transition, such as gas density, metallicity, turbulence, and radiation fields, is essential for linking small-scale cloud physics to the large-scale evolution of galaxies. 
Observations, like those of the Perseus molecular cloud \citep{Lee2015-Perseus_Transition,Park2023}, provide key constraints on these mechanisms, testing theoretical models and revealing the complexity of real interstellar environments.

\begin{figure}[ht]
    \centering
	\includegraphics[width=\columnwidth]{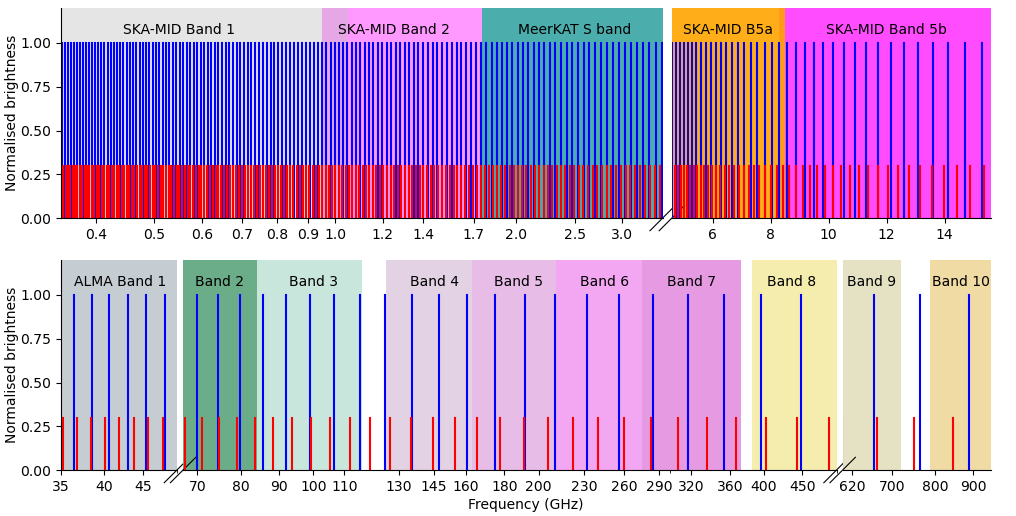}
    \caption{Hydrogen RRLs covered in the frequency bands of SKA telescopes and ALMA (and MeerKAT's S-Band for completeness). For clarity, only the H$n\alpha$ and H$n\beta$ lines are shown. Typically, the H$n\beta$ lines are $\sim$30\% of the brightness of the H$n\alpha$ lines. There are a similar number of H$n\gamma$ and He$n\alpha$ lines in the frequency bands, both of which are $\sim$10\% of the brightness of the H$n\alpha$ lines.}
    \label{fig:rrls}
\end{figure}

Perhaps even more importantly, the combined spectral coverage of the SKA telescopes and ALMA will enable the observation of an unprecedented number of radio recombination lines (RRLs), tracing a wide range of physical conditions. This can be seen in Figure \ref{fig:rrls}, which shows the frequencies and relative brightnesses of the H$n\alpha$ ($\Delta\,n = 1$) and H$n\beta$ ($\Delta\,n = 2$) recombination lines. As the frequency spacing of RRLs becomes much closer at lower frequency, SKA-Mid will be able to simultaneously observe 82 H$n\alpha$ lines in Band 1, 36 in Band 2, and 21 in Band 5a. As lines of H, He, and C are also found close together in frequency,  with sufficient sensitivity, similar numbers of H$n\beta$, H$n\gamma$, He$n\alpha$, and C$n\alpha$ lines can also be observed at the same time. With an instantaneous bandwidth of 2$\times$2.5 GHz SKA-Mid Band 5b cannot be simultaneously observed in its entirety, but nevertheless $\sim$12 H$n\alpha$ lines could be simultaneously observed depending on the precise tuning windows. 

Generally speaking, ALMA will contribute observations of high-frequency RRLs (e.g., H30$\alpha$ at 231~GHz), tracing relatively dense ionised gas, for example, in H\,{\sc ii} regions. This will be complemented by low-frequency RRLs traced by the SKA telescopes, which will be sensitive to more diffuse ionised gas (see also \citealp{Karska01.2026.SKA}). 
Since these low-frequency tracers of ionised gas are virtually unaffected by extinction, the combination of both will improve the identification of H\,{\sc ii} regions embedded in more extended ionised gas. SKA-Mid will also be able to constrain the free-free contribution to the continuum and the line-to-continuum ratio (which is used to derive the electron temperature). The former is particularly important for RRL studies at 3~mm wavelengths, where the continuum emission can arise from a mixture of free-free and dust emission, making it difficult to estimate the true line-to-continuum ratio. Synthetic observations of RRLs toward model H\,{\sc ii} regions with ALMA and EVLA were presented by \citet{pet12}. 
Beyond hydrogen, also carbon recombination lines (e.g., \citealp{cro25}) can be observed across both bands, cm and mm.

Additionally, there are several molecular species with transitions across the radio and submm frequency range. 
Formaldehyde (\ce{H2CO}) provides a diagnostic of density and kinetic temperature in the interstellar medium when combining ALMA and SKAO capabilities. At centimetre wavelengths, SKA-Mid will detect \ce{H2CO} in absorption against continuum sources, enabling density measurements of gas even at low volume densities. In contrast, ALMA accesses the millimetre \ce{H2CO} ladder in emission, tracing warmer and denser molecular gas within star-forming regions and protostellar environments. Together, these regimes provide a wide density baseline from the diffuse ISM to compact cores (e.g., \citealp{man93,gin11,imm16}), enabling direct constraints on molecular gas across scales. Synergistic observations would therefore allow for tomography of gas density structure using a single molecule across up to five orders of magnitude in density, tightly linking global cloud conditions to localised star formation.

Other key molecules with transitions accessible by both SKA telescopes and ALMA include hydroxyl (\ce{OH}), methylidyne (\ce{CH}), and methanol (\ce{CH3OH}), where the cm range accessible to SKA-Mid encompasses maser transitions while ALMA is sensitive to rotational transitions (see also \citealp{Karska01.2026.SKA}). 
For a molecule like HCN, which is perhaps most famous as a 'dense` gas tracer in the ISM (e.g., \citealp{gao04}), $l$-bending transitions can be observed in the cm range. While rotational transitions in the mm range are collisionally excited and depend on density and temperature, transitions within vibrationally excited bending modes trace warm, dense gas near strong infrared sources, where IR pumping populates the bending modes (e.g., \citealp{tho03}).

\subsection{Young Stellar Objects}

The synergy between ALMA and SKA-Mid will allow us to explore the rich chemistry of low-mass young stellar objects. ALMA is particularly effective at detecting transitions of many interstellar complex organic molecules (iCOMs\footnote{iCOMs = interstellar Complex Organic Molecules: C-bearing species with $>$6 atoms and containing heteroatoms as Oxygen and/or Nitrogen \citep{Herbst2009,Ceccarelli2023}.}), such as formamide (\ce{NH2CHO}) and glycolaldehyde (\ce{CH2OHCHO}). The emission peaks for these species, which usually contain elements like carbon, oxygen, nitrogen, and sulfur, are in the spectral range covered by ALMA, especially at warm/hot temperatures ($\geq$100 K).
The observations with SKA-Mid Band 5 complement such ALMA observations in two important ways. 
First, it can detect long carbon chains and rings, such as cyanopolyynes (up to \ce{HC11N}), benzyne (\ce{o-C6H4}), indene (\ce{C9H8}), and glycolamide (\ce{NH2C(O)CH2OH}). These heavy molecules, whose emission peaks at lower frequencies and are therefore not accessible with ALMA, have been detected in cold environments ($<$10 K) with single-dish telescopes \citep{mcguire_gotham_2020, remijan_gotham_2024, Xue_gotham_2025, cernicharo_quijote_2021, Fuentetaja_quijote_2022, rivilla_glycolamide_2023}. 

\begin{figure}[ht]
    \centering
	\includegraphics[width=\columnwidth]{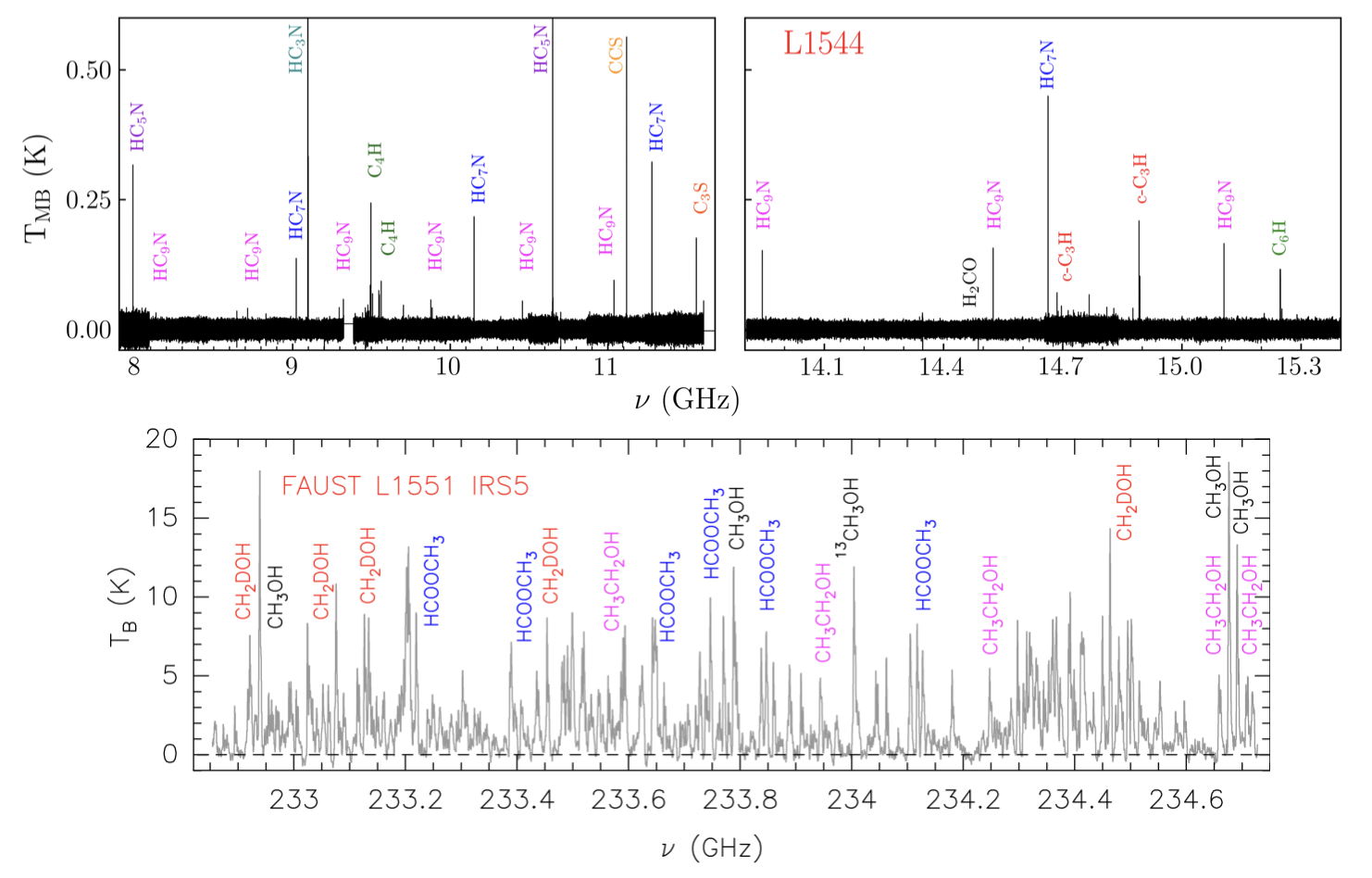}
    \caption{Spectra showcasing complex molecular species detected across two distinct frequency ranges. The low-frequency data toward the prestellar core L1544 (Upper panel) were obtained with the 100 m Robert C. Byrd Green Bank Telescope \citep{Bianchi2023, Giani2025b}. The (sub-)millimetre emission toward the young protostar L1551 IRS5 (lower panel) was obtained with ALMA as part of the FAUST Large Program \citep{Bianchi2020, Codella21}. 
    Complementary radio and (sub-)millimetre observations provide access to different species and distinct excitation conditions, enabling a complete chemical inventory of the earliest stages of star and planet formation.}
    \label{fig:spectra-GBT-ALMA}
\end{figure}

Although the abundances of these species are low (10$^{-12}$ to 10$^{-6}$ with respect to H$_2$; e.g., \citealp{cec23}), they may play a crucial role in transferring organic material from early star-forming phases to bodies in a planetary system, such as asteroids and comets. Therefore, combining ALMA and SKA-Mid will provide a complete picture of the chemical complexity present in the early stages of planetary system formation (Figure \ref{fig:spectra-GBT-ALMA}).
Second, in the inner regions of a protostellar disk, dust emission is often optically thick within the ALMA frequency range \citep{Lee2017, DeSimone2020-COMs_VLA}. This obscures molecular emission in the inner part of the disk and can lead to incorrect measurements of gas physical parameters. Indeed, any radiative transfer analysis that relies on multiple molecular transitions may result in an inaccurate estimation of gas temperature and column density if the line emission is partially obscured by dust \citep[][]{frediani_2025}. This attenuation can also vary between different transitions if they do not originate from the same physical region \citep{Bianchi_siblings_2022}. Observations with the Karl G. Jansky Very Large Array (VLA) of the protostellar binary system NGC1333 IRAS 4A at 25~GHz \citep{des22} have shown that $\tau_{\rm dust}$ is $\gtrsim$1.6 at 143~GHz, enough to attenuate the methanol line intensity by a factor $\gtrsim$5. $\tau_{\rm dust}$ is $\gtrsim$4.9 at 250~GHz and larger at higher frequencies. SKA-Mid will enable the observation of molecular emission in a frequency range where the dust is less optically thick. This will allow for the discovery of complex organic molecular emission in sources where it was previously obscured by dust in ALMA bands, including in more massive disks around high mass protostars. Such SKA-Mid observations will provide crucial constraints on the dust's optical depth, which will help us determine how much dust is attenuating the molecular emission in the ALMA bands and, in turn, correct our determined gas temperatures and column densities. \citet{DeSimone2020-COMs_VLA} demonstrated that at frequencies around 25~GHz the dust can already be optically thin, revealing molecular line emission from inner regions, although this is not the case for all sources. SKA-Mid will probe even lower frequencies, allowing dust opacity effects to be fully characterised and uncovering molecular emission still obscured at ALMA wavelengths. For a broader discussion of dust opacity effects and their implications for chemical complexity, see \citet{Bianchi01.2026.SKA}.

Several species, such as \ce{HC3N} and \ce{HC5N}, provide a powerful probe of interstellar gas properties. 
These molecules have transitions from millimetre to centimetre wavelengths—meaning ALMA can resolve their high-J lines in dense, warm regions, while SKA-Mid accesses their low-J, low-frequency lines that trace cooler, more diffuse, or extended gas. 
Because these species are sensitive to both density and temperature, as well as to chemical timescales, combining data from both facilities allows us to constrain excitation conditions, optical depths, and molecular abundances with far greater precision. 
Moreover, the long carbon chains in \ce{HC3N} and \ce{HC5N} are excellent tracers of early-stage chemistry in star-forming regions \citep[e.g.,][]{Sakai13, Taniguchi2024-,liu20}, and their distribution can reveal the interplay between cold chemistry, warm-up processes, and energetic feedback. The ratio \ce{HC3N}/\ce{HC5N} measured in the protostellar cluster OMC-2 FIR4, has been used to provide important constraints on the cosmic-ray ionisation rate \citep[$\zeta$,][]{Fontani2017}, a key parameter for star formation \citep{Padovani+2009,Padovani2020-CosmicRays_Star_Formation,Kuffmeier2020-Ionization_Disk_Size}.
The ALMA--SKA-Mid synergy enables us to connect the chemistry of small, dense cores to larger-scale cloud structures and produce maps of the cosmic-ray ionisation rate \citep{Sabatini2023-High_mass_Zeta,Pineda2024-ProPStar_NGC1333_CRIR}.

This ability to trace chemistry across different physical scales is particularly crucial when studying energetic feedback mechanisms, such as from jets and molecular outflows, which dominate the large-scale kinematics of star-forming regions (see \citealp{Sabatini01.2026.SKA}). These ejections of material play a crucial role in removing angular momentum from the accreting system, setting the final mass of forming stars \citep[e.g.,][and references therein]{Frank14}. The highly collimated, high-velocity ($\sim$100~km~s$^{-1}$) jet accelerates the surrounding, denser, molecular material, generating a slower ($\sim$10~km~s$^{-1}$) molecular outflow that can extend over distances of $>0.1$~pc \citep[e.g.,][and references therein]{Lee20}. The astrochemical realm revealed by ALMA showed that outflows can be chemically rich in iCOMs such as methanol (CH$_3$OH), acetaldehyde (CH$_3$CHO), methyl formate (CH$_3$OCHO), and formamide (NH$_2$CHO) -- e.g., \cite{joh20,Bonfand24, Lopez-Sepulcre24, Sabatini24b}. Complementary to ALMA's detection of iCOMs at mm-wavelengths, the SKAO will not only allow us to study the chemistry of long carbon chains and rings in molecular outflows across the full range of stellar masses, but will enable us to provide, for the first time, insight into the refractory material of dust grains. 
The low $J$ transitions of heavy, simple, metal-bearing species such as NaCl, 
and other species containing S-, Al-, Mg-,  fall in to the SKA-Mid Band 5 frequency range and may provide valuable probes of regions where grains are forming or are destroyed \citep{2023ApJ...942...66G,1987A&A...183L..10C}, complementing ALMA observations of the higher $J$ transitions. 

Alongside this chemical complexity, ALMA observations are crucial for constraining the physical properties and transport of interstellar dust grains within the envelope-disc systems of YSOs. Growing evidence suggests that dust can begin to grow to mm-sizes much earlier than previously thought \citep[e.g.,][]{Maureira24}, a result that cannot be fully explained by numerical simulations, which predict inefficient dust growth at the typical densities of collapsing protostellar envelopes  \citep[$\sim10^5$~cm$^{-3}$; e.g.,][]{Bate22}. Therefore, it has been proposed that large grains efficiently grown in the inner disc are transported into the envelope via magnetohydrodynamic (MHD) winds \cite[e.g.,][]{Giacalone19, Tsukamoto21, Uchimura25}: a scenario supported by the anti-correlation between the dust spectral index and jet mass-loss rate \citep{Cacciapuoti24}. Observations of the dust opacity spectral index at mm-wavelengths ($\beta_{\rm mm}$) provide strong evidence that outflows and winds are crucial in promoting dust growth and transport. For example, the detection of low $\beta_{\rm mm}\sim1$ values (i.e. $>100$~$\mu$m grains) along outflow cavity walls \citep{Sabatini24b, Sabatini25} -- reported by the ALMA Large Program ``Fifty AU STudy of the chemistry in the disk/envelope system of Solar-like protostars'' \cite[FAUST;][]{Codella21} -- suggests dust is being entrained by outflow activity into the envelope, where comparable $\beta_{\rm mm}$ have been derived  \citep{Cacciapuoti25}. However, accurate $\beta_{\rm mm}$ measurements are often challenging due to contamination from free-free emission. At $\sim$90~GHz, for instance, up to 20\% of the observed flux density may originate from free-free processes \citep[e.g.][]{Hull16} -- or more in the case of high-mass YSOs. Furthermore, as this effect varies over time, accurately quantifying the free-free contribution at mm-wavelengths requires dedicated multi-epoch, multi-frequency observations at lower frequencies. SKA-Mid Band 5a and 5b, combined with ALMA multi-band observations, will provide the required resolution, sensitivity, and frequency coverage to build resolved dust Spectral Energy Distributions (SED) from sub-mm to cm wavelengths. This synergy will allow the definitive study of how dust grows and is transported between the disc and the envelope \citep[e.g.,][]{Tsukamoto21, Morbidelli24, Cacciapuoti25, Sabatini25}.

In the case of high-mass star formation, some of the aforementioned spectral line science also applies to the envelopes and disks of individual massive YSOs \citep[e.g.,][]{Ginsburg19, Colzi21,Sabatini21, Sabatini24a, Wang22,Hoque25}. Key thermal molecular lines that are known to be evolutionary tracers \citep[]
{Sakai2010, Taniguchi2019} can be targeted, including early-phase chemistry tracers (carbon-chains) such as cyanopolyynes (e.g. HC$_3$N) observed with the SKA telescopes, complemented with other tracers observed with ALMA such as deuterated species, shock tracers (e.g., SiO) and quiescent dense gas tracers (e.g., N$_{2}$H$^{+}$). Combined with continuum emission across centimetre to submillimetre wavelengths tracing free-free emission of radio jets/H\,{\sc ii} regions (SKA-Mid) and thermal dust emission from disks/envelopes (ALMA), SKAO--ALMA synergy will contribute to constrain the evolutionary phases of high-mass YSOs.

The host of radio recombination lines (RRLs) available at ALMA and SKAO frequencies, combined with line stacking, will also allow the dynamics of the ionised gas towards massive YSOs to be studied in detail \citep[e.g.,][]{Keto08, Moscadelli21,Sanchez-Monge25}. The line-of-sight motions determined from the line analysis can be combined with time-domain proper motion studies to map the 3D structure of jets and HII regions, revealing how accretion and collimation processes take place in massive star formation. The RRLs may come in the form of maser recombination line emission \citep[e.g.,][]{Jimenez-Serra13, Zhang19} and thermal recombination lines \citep[e.g.,][]{Sanchez-Monge25}. The improvement in sensitivity provided by the SKA-Mid at cm wavelengths will also allow more MYSOs to be studied in this way.

A subject of particular interest for massive star formation is the wider range of maser transitions that the SKAO will make available at high resolution, such as hydroxyl and methanol masers (see also \citealp{Rygl01.2026.SKA}). ALMA will complement these with corresponding high-excitation maser transitions that are relevant in regions of high-mass star formation \citep[e.g.,][]{Hirota21}. Multiple maser transitions of different molecular species have been proposed to trace different dynamical structures and evolutionary stages of high-mass star formation processes \citep[e.g.,][ and references therein]{Urquhart2024}. For example, class I methanol masers (e.g. 44~GHz and 95~GHz) and the 22~GHz water masers mainly trace outflows and jets, the 6.7~GHz class II methanol masers are sometimes associated with disks/envelopes in young high-mass protostellar objects, and the 1.6~GHz OH masers are distributed around H\,{\sc ii} regions in more evolved phases (see \citealp{Rygl01.2026.SKA}). Larger surveys of masers at high resolution with the SKA telescopes and ALMA will improve statistical studies to determine the evolutionary sequence of high-mass YSOs. For example, SKA-Mid can survey the 6.7~GHz methanol and 1.6~GHz OH masers and ALMA can cover some Class I and II methanol masers at higher frequencies \citep[e.g.,][]{Baek23,Zinchenko25}. The monitoring of an unprecedented number of maser transitions toward individual high-mass star formation regions across the radio spectrum will provide new constraints on the interplay of maser variability across different excitation levels. We discuss the prospects of simultaneous measurements and time domain science more generally in the next section. 

\section{Simultaneous measurements and time domain science\label{sec_sim}}

\begin{figure}[ht]
    \centering
	\includegraphics[width=0.35\columnwidth]{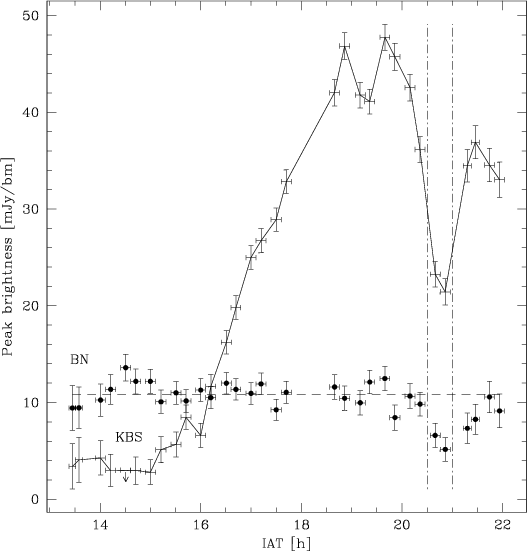}
    \raisebox{15pt}{\includegraphics[width=0.55\columnwidth]{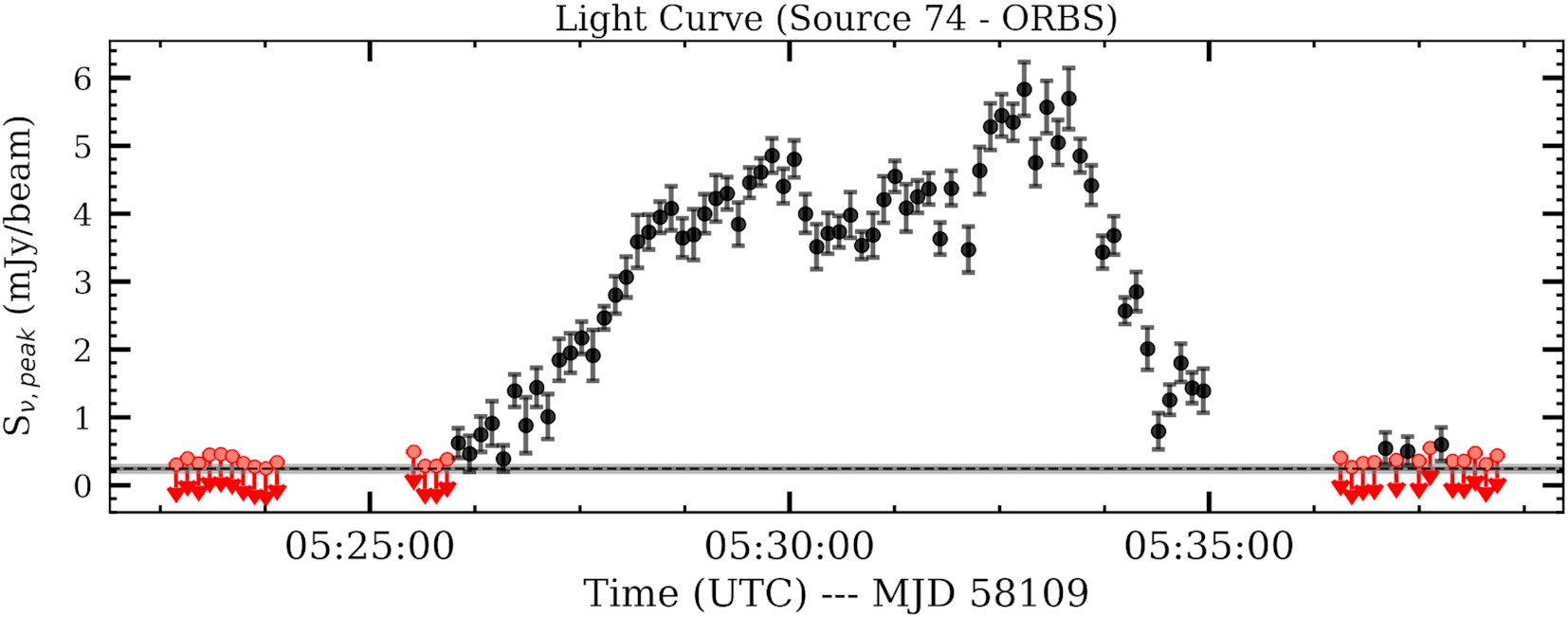}}
    \caption{Radio light curves of the embedded YSO ORBS: {\it left:} a VLA light curve at 22~GHz (from \citealp{for08}), and {\it right:} an ALMA 100~GHz light curve (from \citealp{var23}).}
    \label{fig:ori2}
\end{figure}

With increased wide-band sensitivity, a better signal-to-noise is achieved on shorter timescales, which enables the use of time-sliced imaging to study continuum emission, in particular in the time domain. 
Until recently, and particularly when considering YSOs, extreme variability was only found serendipitously, if across the cm-mm radio range \citep[e.g.,][]{bow03,for08}, the VLA upgrade and the advent of ALMA have already enabled more systematic searches \citep[e.g.,][]{for17,var23}. 
As an example of a deeply embedded YSO in Orion with spectacular flares in both the cm and the mm range (22~GHz and 100~GHz), see Figure~\ref{fig:ori2}. As summarised above, this is non-thermal variability most likely due to (gyro-)synchrotron radiation from the corona of this YSO. To make progress in understanding the origin of this emission and its potential impact in the protostellar system, simultaneous observations covering the cm and mm wavelength domains will be necessary, not least to obtain more reliable spectral index time series (as a poor man's version of solar dynamic spectra) of extreme variability. While such simultaneous observations are beginning to be feasible between the VLA and ALMA, the combination of SKA telescopes and ALMA, with both observatories in the same hemisphere, will be considerably more powerful.

In addition, time-domain studies of centimetre and (sub)millimetre continuum emission will be used to reveal flux variation and proper motion measurements, as they can directly trace new jet ejection \citep{Cesaroni2018} and changes in accretion luminosities \citep{Hunter2017}, respectively. 
These observational studies can be linked to mass ejection and accretion rates during the single accretion burst event (e.g., \citealt{Sabatini01.2026.SKA}). 
Continuum spectral energy distributions from centimetre (SKA-Mid) to submillimetre (ALMA) will be essential to distinguish emission mechanisms (i.e., free-free emission from radio jet and thermal dust emission from disk/envelope), constrain temperature and mass of the regions. 
Compared with currently available radio interferometers \citep{Wright2023}, SKA telescopes will enhance sensitivity and spatial resolution, and hence, it can cover more high-mass protostellar objects at farther distances (e.g., the highest resolution of 30~mas can resolve structures of 30~au and 300~au scales at 1~kpc and 10~kpc sources, respectively, at 15~GHz).  
Such survey and monitoring studies will be useful to directly test theoretical models of high-mass star-formation processes \citep{Meyer2017} by statistically estimating the event rate of accretion bursts and the duration of each event within a human timescale. 

\subsection{Feasibility}

\begin{figure}[ht]
    \centering
	\includegraphics[width=0.9\columnwidth]{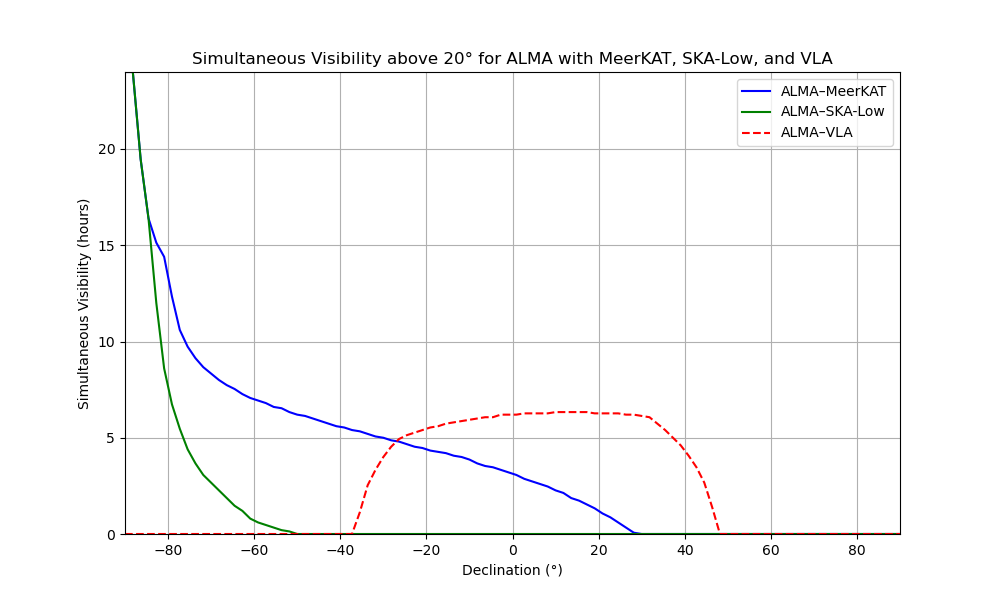}
    \caption{Simultaneous visibility of astronomical objects from the ALMA and SKAO sites, as a function of Declination. The simultaneous visibility between ALMA and VLA is shown for comparison.}
    \label{fig:almaska}
\end{figure}

Simultaneous measurements involving both SKA telescopes and ALMA require joint visibility above the local horizon for a given target. So far, while the joint visibility of a given target between ALMA and the NRAO Very Large Array was enabled by a fairly similar longitude, their different latitudes are currently severely limiting the accessible Declination range. For the combination of ALMA and the SKA telescopes, the latitude difference will no longer be a constraint, in principle opening the entire Southern sky for such observations -- but the considerable longitude difference results in limited hours of simultaneous visibility of a given target. As shown in Figure~\ref{fig:almaska}, the simultaneous visibility of sources for ALMA and MeerKAT is at least as good as that currently accessible to ALMA and VLA at Declinations of $\lnapprox -30^\circ$. 
The sites of SKA-Mid and SKA-Low have only very limited simultaneous visibility.

Apart from sensitivity considerations, the synthesised beam size will be a major constraint for simultaneous monitoring observations of stars. While SKA-Mid will already reach arcsecond beam sizes in early science, the beam size of 5\arcsec for SKA-Low will limit stellar observations to very nearby sources without confusion by background sources. Since ALMA is cycling through different array configurations, optimal matches of suitable beam sizes in given frequency bands between ALMA and the SKA telescopes may be rare occurrences.

These considerations notwithstanding, sensitivity considerations are likely to be a greater concern, together with the accessible field of view. The combination of angular resolution, field of view, and sensitivity significantly favours observations of targets with reasonably predictable variability rather than wide-area combined searches.

\subsection{Maser variability}

Methanol masers at 6.7~GHz around high-mass protostellar objects sometimes show sudden increases in flux densities caused by episodic mass accretion events (e.g., \citealp{Rygl01.2026.SKA}). 
In general, methanol maser flare events have been recognised mainly through collaborative works with single-dish monitoring observation programs \citep{Burns2024}. 
Higher sensitivity SKA-Mid observations with a wider field-of-view will enable catching extreme maser flare events thanks to its commensal observations in the course of Galactic plane survey and other large programs. 
Once the methanol maser flare is identified, follow-up observations with both SKA telescopes and ALMA can be initiated, along with other radio and infrared facilities. 
Variation of centimetre and (sub)millimetre masers/thermal emission lines, such as methanol, will provide information on changes in physical and chemical properties around central YSOs, such as molecular column densities, excitation temperatures, and size of the emission regions \citep{Hirota2022, Burns2023}. 

In the recent accretion burst event in a high-mass protostellar object G358.93-0.03, new methanol maser lines are identified for the first time, both with centimetre \citep{Breen2019} and (sub)millimetre \citep{Brogan2019} interferometer observations. 
Higher sensitivity and wider-band observations with SKA telescopes and ALMA will provide more opportunities to detect rare maser species. 
Multi-transition maser modeling will constrain the physical conditions and their temporal variations in the maser-emitting regions, which are sensitive to hydrogen density, temperature, molecular abundance, and/or radiation field. 

\section{Outlook: SKAO Early Science opportunities and the ALMA2030 Wideband Sensitivity Upgrade}

SKA-Mid early science operations in 2031 will coincide with the increasing availability of new ALMA capabilities enabled by the ALMA2030 Wideband Sensitivity Upgrade (WSU). The most obvious joint ALMA-SKAO synergy in this early period of operations will lie in complementary continuum and spectral line observations 
across the cm and mm wavelength bands, where SKA-Mid in particular will already deliver arcsecond-level angular resolution. 
So far, the VLA has for a long time been the only interferometric facility providing sub-arcsecond resolution at cm-radio wavelengths and high sensitivity, if limited to the northern sky. Southern star-forming regions, such as Corona Australis, therefore lack equivalent radio information, even if ATCA has served as a valuable pathfinder. The SKAO, beginning with its precursors such as MeerKAT, is beginning to fill this gap and will enable the first systematic and coordinated mm- and cm-wavelength observations in the southern hemisphere, making joint studies of nearby low-mass and more distant high-mass star-forming regions possible from the start of SKAO early science. 
Together with ALMA's Wideband Sensitivity Upgrade, the SKAO will offer an unprecedented spectral and spatial view of these sources. The WSU's increased instantaneous bandwidth allows multiple transitions of the same molecular species to be observed in a single setup, enabling more robust derivation of physical and chemical properties through multi-line analysis. The SKA telescopes will extend this coverage into the radio range, helping to separate the effects of dust opacity and free-free emission and providing a complete picture of the emission mechanisms and physical conditions in young stellar environments.
Initial simultaneous ALMA and SKAO observations will become feasible as soon as SKAO early science capabilities come online, opening the path to coordinated multiwavelength investigations of star formation and the interstellar medium. Additionally, nearby star-forming regions, such as Ophiuchus, Orion, and Corona Australis, already possess a wealth of complementary data within the ALMA archive. This existing high-resolution (sub-)millimetre data can be leveraged to provide a comprehensive, multi-wavelength view of the chemical and physical structures of these regions.

The synergy between ALMA and SKAO will provide an unprecedented view of the cold and warm Universe, from cm to mm wavelengths. To fully exploit this potential, extending the SKA-Mid frequency coverage toward higher frequencies would be highly beneficial. Currently, SKA-Mid Band 5 reaches up to 15 GHz, while ALMA Band 1 starts at about 35 GHz, leaving a significant gap between 15 GHz and 35 GHz. The proposed SKA-Mid Band~6 (15–50 GHz) would bridge this gap and open access to a range of key spectral diagnostics that are rarely observable elsewhere and not at the expected sensitivity (see SKA Memo 20-01).
This frequency range includes, for example, the inversion transitions of ammonia (NH$_3$), a powerful tracer of dense and cold gas, and the bright methanol (CH$_3$OH) K-ladders, which have recently proven to be excellent probes of physical and chemical conditions in young protostellar sources (see Section \ref{sec:SpecLine}). 
At present, the VLA is the only interferometer capable of covering this range, but it is restricted to the northern sky and is already used at the maximum of its capabilities. The implementation of SKA-Mid Band 6 would therefore be crucial to bridge this gap and extend these capabilities to the southern hemisphere, allowing comprehensive studies of molecular complexity and the interstellar medium across both low- and high-mass protostars.

\bibliographystyle{abbrvnat-maxbibnames4}
\bibliography{chapter} 

\end{document}